\documentclass[a4paper,11pt]{article}

\usepackage{pos}
\usepackage{lipsum} %package to generate placeholder text in the following
\usepackage{siunitx}
\DeclareSIUnit{\ev}{eV}
\usepackage{graphicx}
\usepackage{xcolor}
\usepackage{wrapfig}
\usepackage{subfigure}

\title{Seasonal Variations of the Atmospheric Neutrino Flux measured in IceCube}

\ShortTitle{Seasonal Variations of the Atmospheric Neutrino Flux measured in IceCube}

\author{The IceCube Collaboration \\{\normalsize \normalfont(a complete list of authors can be found at the end of the proceedings)}\\}

\emailAdd{karolin.hymon@tu-dortmund.de}
\abstract{

The IceCube Neutrino Observatory measures high energy atmospheric neutrinos with high statistics. These atmospheric neutrinos are produced in cosmic ray interactions in the atmosphere, mainly by the decay of pions and kaons. The rate of the measured neutrinos is affected by seasonal temperature variations in the stratosphere, which are expected to increase with the energy of the particle. In this contribution, seasonal energy spectra are obtained using a novel spectrum unfolding approach, the \textit{Dortmund Spectrum Estimation Algorithm} (DSEA+), in which the energy distribution from \SI{125}{\giga\ev} to \SI{10}{\tera\ev} is estimated from measured quantities with machine learning algorithms. The seasonal spectral difference to the annual average flux will be discussed based on preliminary results from IceCube’s atmospheric muon neutrino data.

\vspace{4mm}
{\bfseries Corresponding authors:}
Karolin Hymon$^{1*}$, Tim Ruhe$^{1}$\\
{$^{1}$ \itshape Astroparticle Physics WG Rhode, TU Dortmund University, Germany}\\
$^*$ Presenter

\ConferenceLogo{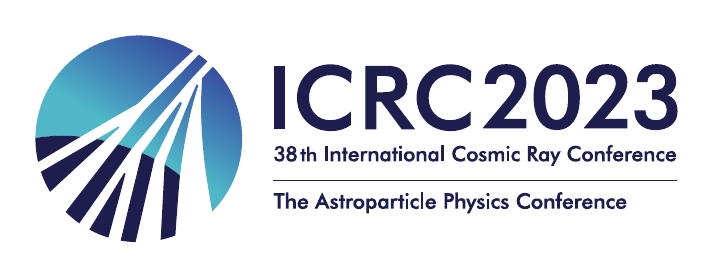}

\FullConference{The 38th International Cosmic Ray Conference (ICRC2023)\\ 26 July -- 3 August, 2023\\ Nagoya, Japan}
}

\begin{document}

\maketitle

\section{Introduction}\label{chapter:seasonalvariations}

% oroginal sentence from last proceedings
%IceCube is a cubic-kilometer neutrino detector installed in the ice at the geographic South Pole between depths of \SI{1450}{\metre} and \SI{2450}{\metre}, completed in 2010. Reconstruction of the direction, energy and flavor of the neutrinos relies on the optical detection of Cherenkov radiation emitted by charged particles produced in the interactions of neutrinos in the surrounding ice or the nearby bedrock \cite{Aartsen:2016}.

% brief intro into why seasonal variations are importan
%\color{red} \textit{here: state why sv are important in general} \color{black}

For current neutrino telescopes, lepton fluxes originating from cosmic ray interactions in the atmosphere impose a challenging background in identifying scarce astrophysical neutrinos.
Besides the requirement of an accurate understanding of leptonic fluxes for detector calibration, they serve as indicators of probing hadronic interactions within cosmic ray-induced particle showers in the atmosphere.

%\color{red} \textit{Where do variations come from? -explain briefly and refer to other proceedings/Jakobs paper?} \color{black}

A hadronic cascade of mesons, mainly composed of kaons and pions, is initiated by an interaction of a primary cosmic ray in the upper atmosphere \cite{gaisserbook}. Conventional atmospheric muons and neutrinos are produced in weak decays by these parent mesons. The relative probability for their decay or re-interaction inside the atmosphere is at equilibrium at the critical energy for each individual meson. The critical energy at a given atmospheric depth is proportional to the local atmospheric air density, which changes anti-proportionally with temperature. The probability of decay increases with temperature so that the muon and neutrino flux becomes temperature-dependent. The flux modulation based on temperature evaluation throughout the year is referred to as seasonal variations. A more detailed overview on the formalism can be found in \cite{gaisserbook, jakobpaper}.

%When a cosmic ray interacts in the upper atmosphere,  a hadronic cascade of  mesons, mainly composed of kaons and pions is initiated. The relative probability for their direct decay or re-interaction before decaying into muons and neutrinos is at equilibrium at the critical energy of the mesons. This critical energy at given atmospheric depth is proportional to the atmospheric density and hence, antiproportional to the atmospheric temperature. The local air density changes with temperature in the stratosphere. Increasing temperatures yield an expansion of the atmosphere which results in %dgive nby ideal gas law

%\color{red} \textit{here: mention what was done previously and what has been improved/what is new} \color{black}

The modulation of the seasonal muon and neutrino rate has been studied extensively with respect to atmospheric temperature in various experiments, e.\ g.\ \cite{MACRO:1997, tilav09, Heix19,jakobpaper}, but no energy-dependent measurement of the spectral shape has been conducted on a seasonal basis. In this contribution, we present an update on the progenitor analysis from \cite{hymon21} to \num{11.5} years of neutrino data from IceCube providing a measurement of the seasonal muon neutrino flux variations with respect to energy. % too vague or say that systematics have been updated?

\section{Neutrino Data}

%\begin{wrapfigure}{r}{0.3\textwidth}
%\vspace{-20pt}
%  \begin{center}
%    \includegraphics[width=0.3\textwidth]{Earth_zoomed.png}
% \end{center}
%  \vspace{-20pt}%10
%  \caption[]{Illustration of the atmospheric region in which the neutrinos are produced, marked in red. The zenith range from \SIrange{90}{120}{\degree} corresponds to geographic latitudes from \SIrange{90}{30}{\degree}S. %A deviation up to approx 10\% is observable between \SI{100}{\giga\ev} up to \SI{10}{\tera\ev}.
%   \label{fig:earth}}
%  \vspace{-5pt}%this was wihtout - before
%\end{wrapfigure}
The IceCube Neutrino Observatory is a neutrino detector in the Antarctic ice at the geographic South Pole, located between depths of \SI{1450}{\metre} and \SI{2450}{\metre} with a comprised instrumented volume of a cubic kilometer. The detector array consists of \num{86} vertical cable strings which are equipped with \num{5160} digital optical modules (DOMs) in total. Energy, flavor, and directional reconstruction of the neutrinos rely on the optical detection of initiated Cherenkov radiation by charged particles produced in the interactions of neutrinos in the surrounding ice or the nearby bedrock \cite{Aartsen:2016}. 
% \textit{describe upgoing and downgoing in terms of event selection \& support by figure} \color{black}

The data sample consists of well reconstructed up-going tracks, classified as  muons induced by neutrino interactions inside the ice, with a purity of more than  \num{99.7}\% \cite{Aachen6yrs}. Down-going neutrinos, produced in cosmic ray induced air showers vertically above Antarctica, are excluded in this analysis as these events are hardly distinguishable within the dominating background of atmospheric muons from the same showers. The zenith arrival direction, $\theta$, of the neutrinos is restricted to the Southern Hemisphere within \SIrange{90}{120}{\degree} below the Tropic of Capricorn. Seasonal temperature variations at production heights of the neutrinos are required for this analysis. The Northern Hemisphere is excluded. A deeper description of the zenith direction and temperature variation throughout the year is given in \cite{Heix19}. The analysis includes events  from May 2011 to December 2022, in which the full detector configuration was complete. This results in 523736 neutrino events within \num{11.3} years of effective livetime. %\color{red} An illustration of the atmospheric pressure layers  in which the neutrino interaction occur is illustrated in Fig.\ \ref{fig:temperature}. \color{black}
%\section{Listing some References}\label{sec2}

%As we discussed in \textsection\ref{sec1}, etc. This is a paper from a previous ICRC \cite{Zoll:2015wcu}. This is a second paper from a previous ICRC \cite{Peiffer:2017vsm}. This is a paper from the current ICRC \cite{Author:2023icrc}.
%Here is an IceCube journal paper \cite{Aartsen:2016nxy} and an external journal paper \cite{Waxman:1998yy}.

\section{Spectrum Unfolding}

The determination of the neutrino energy is an inverse problem as it needs to be inferred from an energy proxy of the induced muon in the ice. The muon, however, is exposed to stochastic energy losses and ionization, which in turn smears out the energy resolution. This inverse problem is addressed by a  spectrum unfolding technique with an incorporated machine learning algorithm. The \textit{Dortmund Spectrum Estimation Algorithm} (DSEA+) \cite{dsea} estimates the probability of a given event corresponding to one of the predefined energy bins in terms of a multinomial classification task, which is performed in an iterative manner. The energy is estimated from two proxies, the number of DOMs triggered in each event and an energy reconstruction for the induced muon track \cite{etrun13}. For further explanation of DSEA+  and the optimized internal parameters refer to \cite{hymon21}.
In contrast to the previous work described in \cite{hymon21}, the unfolding algorithm is trained on Monte Carlo simulation weighted to the neutrino flux calculated by MCEq \cite{mceq}. MCEq solves the cascade equations of particle interactions in  atmospheric cosmic ray air showers numerically, based on different theoretical models. The H3a model \cite{h3a} is selected as primary composition model of the cosmic rays, SIBYLL2.3c \cite{sibyll} as hadronic interaction model, and NRLMSISE-00 \cite{msis} as the empirical atmospheric model. DSEA+ is trained on the annual average predicted flux. No seasonal information is fed into the algorithm so that the seasonal variations strength can be determined independently of  prior assumptions on the expected variation strength. The algorithm is robust against changes in spectral shape compared to its training spectrum so that an observation of deviations in spectral shape caused by seasonal variations is feasible.

The effect of systematic uncertainties on the unfolded spectrum is estimated from simulations with varied detector settings. Pseudo-samples are unfolded and the deviation to the reference unfolded pseudo-sample of average systematic parameters is evaluated for upper and lower constraints on each effect, elaborated below. All associated uncertainties are combined in the quadratic sum for each of the individual positive and negative deviations from the reference unfolded spectrum \cite{hymon21}.
The efficiency of the optical modules in IceCube, the absorption and scattering effects in the glacial ice, and the optical properties of the re-frozen ice of the borehole around the strings are estimated from Monte Carlo simulation with varied respective parameter. %\color{red} Absorption and scattering coefficient are varied separately by $\pm 5$\%, the parameter defining the zenith-dependence of the hole-ice by  $\pm 1$\color{black}. 
The uncertainty of the imposed neutrino flux from MCEq in the weighting of the training sample can be constrained by the propagation of uncertainties from the primary cosmic ray composition and hadronic interaction models, as described in \cite{crflux}. The propagated uncertainties of the neutrino flux are linearly interpolated between \SI{100}{\giga\ev} and \SI{10}{\tera\ev}. Two additional pseudo-samples are weighted to the upper and lower limit of the neutrino flux uncertainty and unfolded. The statistical uncertainty is determined by bootstrapping \cite{bootstrap}, in which the events from the seasonal data sets are sampled with replacement in \num{2000} trials and the standard deviation is calculated for the unfolded number of events in each energy bin. To convert the unfolded event spectrum into a differential flux, the effective area needs to be obtained from Monte Carlo simulations of the detector response and the spectrum is accounted for livetime of the seasonal data sets and solid angle of the arrival directions.

\vspace{2cm}

%\color{red} Nugen erwähnen? 

\section{Results}

\begin{figure}[htp]
\centering
    \subfigure[Zenith range: \SIrange{90}{120}{\degree}]{\includegraphics[width=0.8\textwidth]{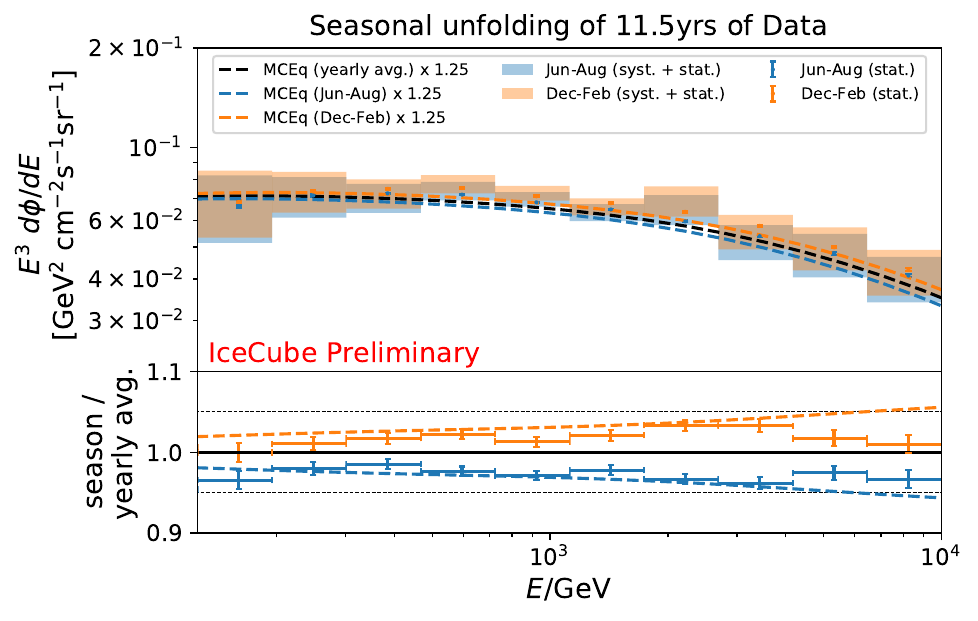}\label{fig:unblind_90-120}}
    \subfigure[Zenith range: \SIrange{90}{110}{\degree}]{\includegraphics[width=0.8\textwidth]{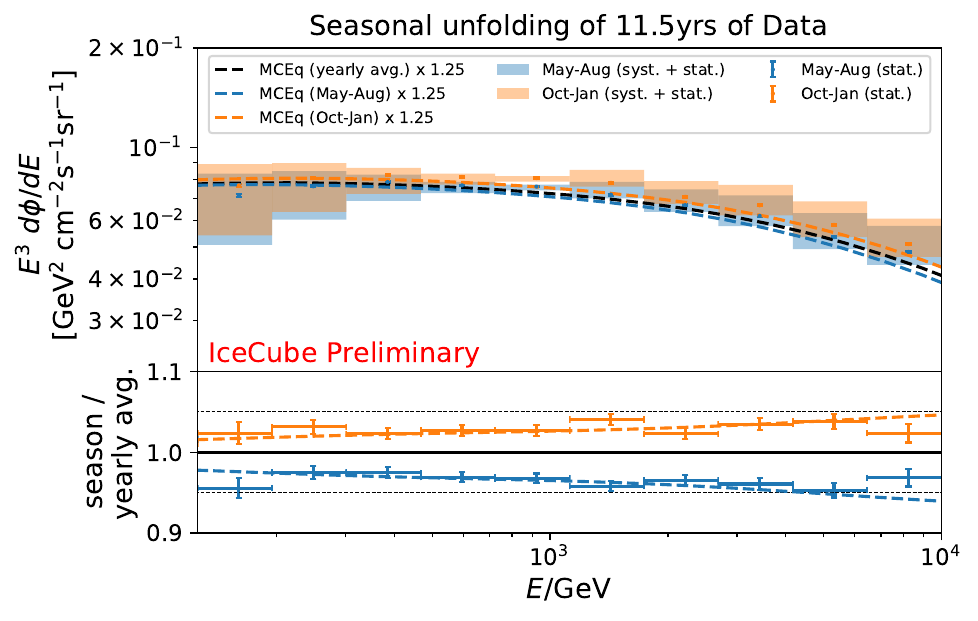}\label{fig:unblind_90-110}}
\caption{Upper panel: unfolded seasonal muon neutrino spectra for austral summer and winter obtained from 11.5 years of IceCube data in the zenith range from \SIrange{90}{120}{\degree} (a) and from \SIrange{90}{110}{\degree} (b). Error bars denote statistical uncertainties, the bands the corresponding systematic uncertainty. The respective predicted flux from MCEq is shown in dashed lines with fitted normalization. Lower panel: ratio of the seasonal to annual average flux for both, the unfolded data and MCEq predictions. Flux deviations of $\pm 5$\% are marked as black dashed lines. Systematic uncertainties remain the same for each season and cancel out in the ratio. \label{fig:unblinded_spectra}} 
\end{figure}

The seasonal spectra are unfolded in ten equidistant bins in $\log(E_{\nu})$ from \SI{125}{\giga\ev} to \SI{10}{\tera\ev} and the variation strength is determined with respect to the annual average neutrino flux. The seasonal flux ratio is merely affected by propagation of the statistical uncertainties of the unfolded rates because the systematic uncertainties are largely independent of the season. %They remain the same throughout the year, so do effective area and solid angle. 
Fig.\ \ref{fig:unblinded_spectra} displays the unfolded seasonal fluxes for two different zenith regions, explained below. The upper panel in each figure shows the unfolded seasonal spectra scaled to $E^3$. The corresponding systematic uncertainties are shown in the shaded bands. The calculated fluxes from MCEq (with H3a, SIBYLL2.3c, NRLMSISE-00) are shown for comparison and are scaled up to match the normalization of the unfolded spectra. The scaling factor is determined by a fit to the unfolded spectrum with respect to the systematic and statistical uncertainties. The loss function considering the asymmetric error bars of the unfolded spectrum is minimized by the Nelder-Mead \cite{Neldermead} approach and a scaling factor of 1.25 is obtained as best-fit value. The lower panel in each figure displays the ratio of the seasonal fluxes to the annual average flux. The ratio displays only statistical uncertainties because the systematic uncertainties cancel out in the ratio as described above. This allows to observe the increase in variation strength with increasing energy at per cent level. 

Fig.\ \ref{fig:unblind_90-120} depicts the unfolded spectra for austral summer (December to February) and winter (June to August) within zenith angles from \SIrange{90}{120}{\degree}.
The unfolded fluxes are in agreement in shape with the seasonal MCEq fluxes. The energy spectrum is flat in the first energy bin due to threshold effects and declines with increasing energy.
The observed seasonal variations increase with the neutrino energy up to \SI{4}{\tera\ev}, as predicted by MCEq. The increase in seasonal variation strength with increasing particle energy is attributed to the interactions of the secondary mesons in the upper stratosphere. High-energy primary cosmic rays interact higher in the atmosphere and produce  secondaries at higher altitude, where the temperature variation throughout the year is larger. 
However, the observed variations decrease above \SI{4}{\tera\ev}, which is not consistent with the MCEq prediction.

\begin{figure}[tbh]
    \centering
    \includegraphics[width=\textwidth]{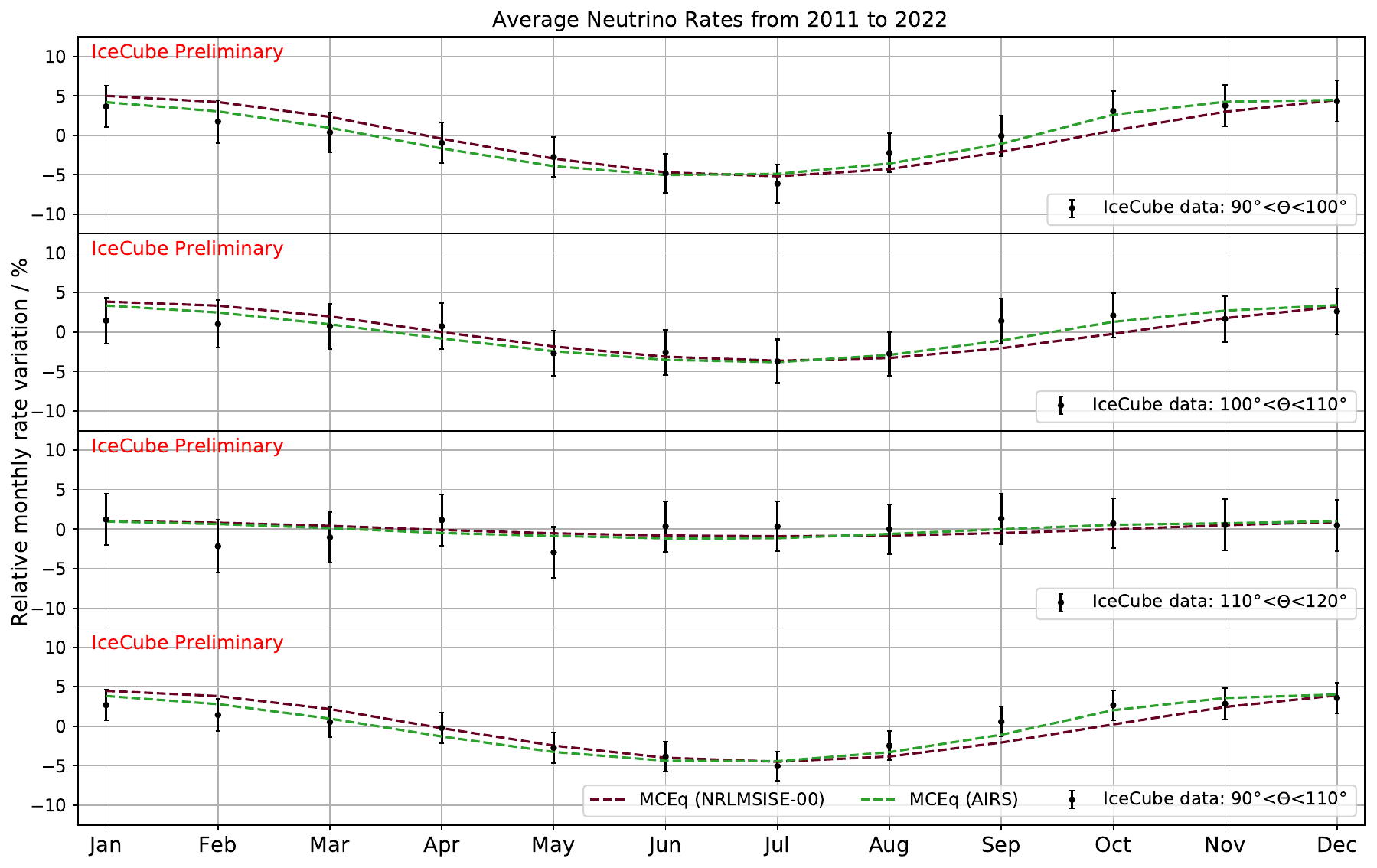}
    \caption{Relative average neutrino rate modulation per month compared to yearly average for the zenith ranges between \SIrange{90}{100}{\degree}, \SIrange{100}{110}{\degree}, \SIrange{110}{120}{\degree} and \SIrange{90}{110}{\degree} including statistical uncertainties. Predicted variations from MCEq are depicted in dashed lines for the atmospheric model NRLMSISE-00 and for five available years of temperature data from the AIRS instrument. }
    \label{fig:mceq_rates}
\end{figure}

To investigate the seasonal variation measurement and the observed decline at energies above \SI{4}{\tera\ev} in detail, the monthly average neutrino rate is investigated in zenith bands of $\Delta \theta =$ \SI{10}{\degree} width. The three upper panels in Fig.\ \ref{fig:mceq_rates} depict the relative average monthly neutrino rate variation with respect to the annual average in the respective zenith bands. The statistical uncertainty is depicted as error bars. The predicted variations from MCEq (H3a, SIBYLL2.3c)  are shown in dashed lines based on the assumption of two different atmospheres. Besides the prediction using the atmospheric model NRLMSISE-00, as for the algorithm training and comparison to the unfolded spectra, a data-based prediction is obtained from temperature data from the Atmospheric Infrared Sounder (AIRS) \cite{airs} on NASA's Aqua satellite. The satellite orbits the Earth twice per day and the AIRS instrument  provides a temperature measurement at pressure levels from \SIrange{0.1}{1000}{\hecto\pascal} with an angular resolution of \SI{1}{\degree} $\times$ \SI{1}{\degree} per longitude and latitude. 
The predicted neutrino rate from MCEq is calculated with daily temperature data from April 2012 - April 2017, as in \cite{jakobpaper}, and the obtained daily rate is averaged per month. The first upper panel in Fig.\ \ref{fig:mceq_rates} shows the rate variation between zenith angles from \SIrange{90}{100}{\degree}. The maximum seasonal variation strength of $(4 \pm 2)$\% is observable in December and January, the minimum of $(6 \pm 2)$\% in July. The variation strength linearly decreases from January to July, a steep increase is observable from July to October. The predicted variations from MCEq are  both compatible with the observed neutrino rate within the statistical uncertainties. The predicted variation strength amplitude using AIRS data is approx.\ \num{1}\% to \num{2}\% higher compared to the calculated prediction with the atmospheric model NRLMSISE-00, except  for June and July. The second panel depicts the relative rate variation between \SIrange{100}{110}{\degree}.  The variation strength decreases over all months compared to the previous zenith range. The rate variation reaches its maximum of approx.\ $(2.5 \pm 2.5)$\%  in December, the minimum with $(4 \pm 2.5)$\% in July. The variation remains roughly constant from January to April and decreases from May to August. The third panel displays the zenith region from \SIrange{110}{120}{\degree}. No variation is evident in the observed neutrino rate. The predicted variations are comparable for both atmospheric assumptions and fluctuate around $0$\%.

\begin{wrapfigure}{r}{0.5\textwidth}[tbh]
\vspace{-20pt}
  \begin{center}
    \includegraphics[width=0.5\textwidth]{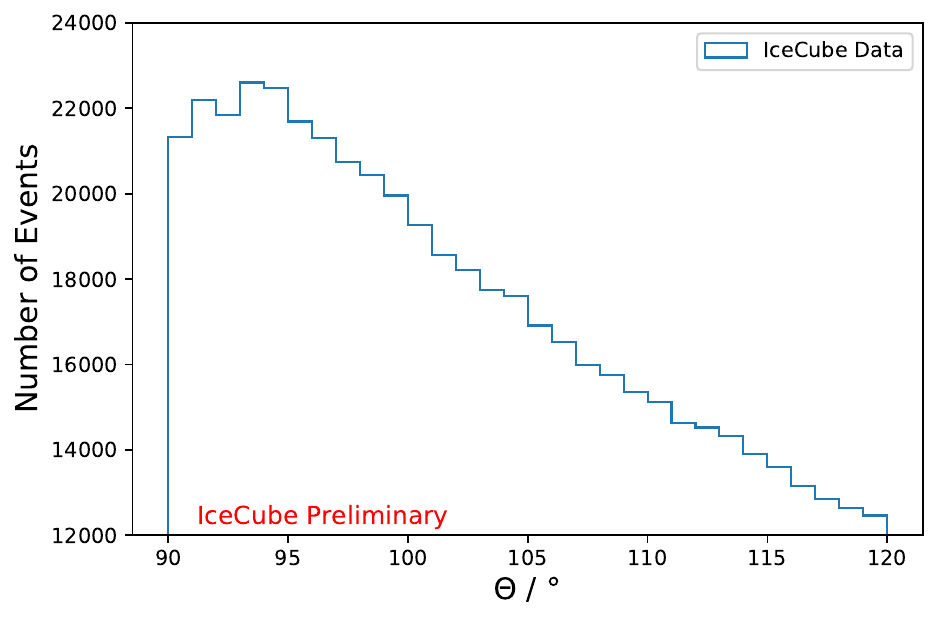}
 \end{center}
  \vspace{-20pt}%10
  \caption[]{Zenith distribution of the neutrino sample.} %A deviation up to approx 10\% is observable between \SI{100}{\giga\ev} up to \SI{10}{\tera\ev}.
  \label{fig:zenith}
  %\vspace{-15pt}%this was wihtout - before
\end{wrapfigure}

Since no seasonal variations are observed within the third zenith band, a deeper investigation of the seasonal modulation is conducted by removing events arriving from zenith angles between \SIrange{110}{120}{\degree}. The zenith range from \SIrange{90}{110}{\degree} is depicted in the lower panel in Fig.\ \ref{fig:mceq_rates}. 
The maximum relative rate variation is in December with $(4 \pm 2)$\%, the minimum in July with $(5 \pm 2)$\%. The variation strength decreases smoothly and slowly from January to July, which is attributed to radiation cooling in the atmosphere, whereas a steep increase in the rate from July to October is likely originating from rapid warming of the stratosphere during sunrise in the Southern Hemisphere. A similar modulation pattern is observed in the muon rate measured in IceCube and its progenitor detector AMANDA \cite{tilav09}. The observed rate variation is similar to the rate variation in the zenith range between \SIrange{90}{100}{\degree} in the upper panel in Fig.\ \ref{fig:mceq_rates}. It indicates that the seasonal modulation within \SIrange{90}{110}{\degree} is dominated by events close to the horizon. The zenith distribution of the neutrino sample is shown in \ref{fig:zenith}.  The zenith cut at $\Theta = \SI{110}{\degree}$ results in a reduction of 26\% of neutrino events in the same detector uptime. 

Fig.\ \ref{fig:unblind_90-110} displays the seasonal unfolding for the zenith range from \SIrange{90}{110}{\degree}.The definition of austral summer (October to January) and winter (May to August) are adjusted compared to Fig.\ \ref{fig:unblind_90-120} by the selection of months with similar average monthly neutrino rates,  observable in Fig.\ \ref{fig:mceq_rates}. The unfolded seasonal event samples of both zenith bands have comparable statistics  so that the reduction of the zenith region does not impact the measurement sensitivity. The unfolded energy spectra are comparable to the spectra in Fig.\ \ref{fig:unblind_90-120} with the same fitted scaling factor for the MCEq normalization. The ratio of the unfolded seasonal flux to the annual average increases with energy from approx.\ $(2\pm 1$)\% at \SI{125}{\giga\ev} up to $(4 \pm 1)$\% at \SI{7}{\tera\ev} for October to January. Despite a drop in the ratio to approx.\ $(4.5 \pm 1)$\% in the first energy bin, the ratio decreases up to \SI{7}{\tera\ev} for May to August, as expected from MCEq. A decrease in the seasonal variation strength with an absolute value of approx.\ $2.5$\% is apparent in the highest energy bin between \SIrange{7}{10}{\tera\ev} for both austral summer and winter.

%\begin{figure}[tbh]
%    \centering
%    \includegraphics[width=0.99\textwidth]{unblinded_e3_12yr_may-aug_oct-jan_sys_90-110_ratio_cblind.pdf}
%    \caption{Upper panel depicts the unfolded seasonal muon neutrino spectra for the updated seasons of austral summer and winter obtained from 11.5 years of IceCube data in the zenith range from \SIrange{90}{110}{\degree}. Errorbars denote statistical uncertainties, the bands the corresponding systematic uncertainty. The respective predicted flux from MCEq is shown in dashed lines, with fitted scaling factor. Lower panel depicts the ratio of the seasonal to annual average flux for both, the unfolded data and MCEq predictions. Flux deviations of $\pm 5$\% are marked as black dashed lines. Systematic uncertainties remain the same for each season and cancel out in the ratio. }
%    \label{fig:unfolding_90-110}
%\end{figure}

\section{Conclusion}

This work presents the first measurement of seasonal  atmospheric muon neutrinos spectra in the zenith range from \SIrange{90}{120}{\degree} and \SIrange{90}{110}{\degree}. The zenith region from \SIrange{110}{120}{\degree} is removed for further investigation as no seasonal modulation is observable. The unfolded modulation of the austral summer and winter relative flux variation for neutrinos from the zenith region between \SIrange{90}{110}{\degree} increases with the neutrino energy up to \SI{7}{\tera\ev}. The observed pattern is expected, as the interactions of the parent meson becomes the dominant process with increasing energy compared to instant decay. These interactions occur at higher altitudes in the stratosphere, which  have a stronger relative temperature modulation throughout the year than lower altitudes. However, the observed variation strength decreases between \SIrange{7}{10}{\tera\ev} despite the expectation from MCEq. 
%The analysis of the two zenith regions both show a trend on declined seasonal variation strength at several \si{\tera\ev}. A potential reason for this could be the atmospheric model NRLMSISE-00 used for the MCEq flux calculation. An approach in incorporating temperature data from satellite measurements instead of an atmospheric model might provide a more realistic seasonal flux scenario. 
Further investigations are required to determine the cause of the decline. The prompt component of the atmospheric neutrino flux, defined as neutrinos produced by prompt decays from charmed mesons, has not been observed so far. This component is seasonally-independent due to the instant decay of the parent particle. However, this component is expected to have a dominant contribution to the atmospheric neutrino flux only above \SI{100}{\tera\ev} \cite{gaisserbook}, so it is unlikely that the decrease in variation strength at few \si{\tera\ev} is attributed to prompt neutrinos. Furthermore, the seasonal modulation at high energies can be investigated by analyzing sub-samples defining a finer seasonal grid throughout the year.
The detector enhancement to IceCube-Gen2 \cite{gen2} and the collection of additional data will constrain the measurement of seasonal variations even further while reducing the statistical uncertainty. An extension  of the energy range beyond \SI{10}{\tera\ev} would give conclusion whether the trend of decreasing  seasonal variation strength persists due the increasing contribution of prompt neutrinos to the conventional atmospheric flux. %with increasing neutrino energy.  %Next-generation reconstruction algorithms the analysis to energies above \SI{10}{\tera\ev} in order to determine whether the trend of decreasing variation strength persists with increasing neutrino energy.

% Bibtex references:
\bibliographystyle{ICRC}
\bibliography{references}

\providecommand{\href}[2]{#2}\begingroup\raggedright\begin{thebibliography}{10}

\bibitem{gaisserbook}
T.~K. Gaisser, R.~Engel, and E.~Resconi,
  \href{http://dx.doi.org/10.1017/CBO9781139192194}{{\em Cosmic Rays and
  Particle Physics}}.
\newblock Cambridge University Press, 2~ed., 2016.

\bibitem{jakobpaper}
{\bfseries IceCube} Collaboration, R.~Abbasi {\em et~al.}
  \href{https://arxiv.org/abs/2303.04682}{arXiv:2303.04682}.

\bibitem{MACRO:1997}
{\bfseries MACRO} Collaboration, M.~Ambrosio {\em et~al.}
  \href{http://dx.doi.org/10.1016/S0927-6505(97)00011-X}{{\em Astropart. Phys.}
  {\bfseries 7} (1997) 109--124}.

\bibitem{tilav09}
{\bfseries IceCube} Collaboration {\em PoS} {\bfseries ICRC2009} (2010) 10.

\bibitem{Heix19}
{\bfseries IceCube} Collaboration {\em PoS} {\bfseries ICRC2019} (2020) 465.

\bibitem{hymon21}
{\bfseries IceCube} Collaboration
  \href{http://dx.doi.org//10.22323/1.395.1159}{{\em PoS} {\bfseries ICRC2021}
  (2021) 1159}.

\bibitem{Aartsen:2016}
{\bfseries IceCube} Collaboration, M.~G. Aartsen {\em et~al.}
  \href{http://dx.doi.org/10.1088/1748-0221/12/03/P03012}{{\em JINST}
  {\bfseries 12} no.~03, (2017) P03012}.

\bibitem{Aachen6yrs}
{\bfseries IceCube} Collaboration, M.~G. Aartsen {\em et~al.}
  \href{http://dx.doi.org/10.3847/0004-637X/833/1/3}{{\em ApJ} {\bfseries 833}
  (2016) 3}.

\bibitem{dsea}
M.~Bunse {\em https://sfb876.tu-dortmund.de/deconvolution/index.html} .

\bibitem{etrun13}
{\bfseries IceCube} Collaboration, R.~Abbasi {\em et~al.}
  \href{http://dx.doi.org/10.1016/j.nima.2012.11.081}{{\em Nucl. Instrum. Meth.
  A} {\bfseries 703} (2013) 190--198}.

\bibitem{mceq}
A.~Fedynitch, R.~Engel, T.~K. Gaisser, F.~Riehn, and T.~Stanev
  \href{http://dx.doi.org/10.1051/epjconf/20159908001}{{\em EPJ Web Conf.}
  {\bfseries 99} (2015) 08001}.

\bibitem{h3a}
T.~K. Gaisser {\em Astroparticle Physics} {\bfseries 35} (2011) 801--806.

\bibitem{sibyll}
F.~Riehn, R.~Engel, A.~Fedynitch, T.~K. Gaisser, and T.~Stanev
  \href{http://dx.doi.org/10.22323/1.236.0558}{{\em PoS} {\bfseries ICRC2015}
  (2016) 558}.

\bibitem{msis}
J.~Picone, A.~Hedin, D.~Drob, and A.~Aikin
  \href{http://dx.doi.org/10.1029/2002JA009430}{{\em Journal of Geophysical
  Research} {\bfseries 107} (12, 2002) }.

\bibitem{crflux}
A.~Fedynitch, J.~Becker~Tjus, and P.~Desiati
  \href{http://dx.doi.org/10.1103/PhysRevD.86.114024}{{\em Phys. Rev. D}
  {\bfseries 86} (Dec, 2012) 114024}.

\bibitem{bootstrap}
B.~Efron and R.~Tibshirani {\em Statistical Science} {\bfseries 1} (1986)
  54--75.

\bibitem{Neldermead}
J.~A. Nelder and R.~Mead {\em Computer Journal} {\bfseries 7} (1965) 308--313.

\bibitem{airs}
I.~S.~T. Teixeira, ``{IRS/Aqua L3 Daily Standard Physical Retrieval (AIRS-only)
  1 degree x 1 degree V006, Greenbelt, MD, USA, Goddard Earth Sciences Data and
  Information Services Center (GES DISC)}.''
\newblock \url{https://acdisc.gesdisc.eosdis.
  nasa.gov/data/Aqua_AIRS_Level3/AIRS3STD.006}.

\bibitem{gen2}
{\bfseries IceCube} Collaboration, M.~G. Aartsen {\em et~al.}
  \href{http://dx.doi.org/10.1088/1361-6471/abbd48}{{\em Journal of Physics G:
  Nuclear and Particle Physics} {\bfseries 48} (06, 2021) 060501}.

\end{thebibliography}\endgroup

% Alternatively, you can include references by hand:
%\begin{thebibliography}{99}
%\bibitem{...}
%
%\end{thebibliography}

\clearpage

%The following list of authors, affiliations and funding agencies will be updated at the day of submission. The following template is a placeholder generated via https://authorlist.icecube.wisc.edu/icecube on March 25, 2023 and will be updated.
\section*{Full Author List: IceCube Collaboration}

\scriptsize
\noindent
R. Abbasi$^{17}$,
M. Ackermann$^{63}$,
J. Adams$^{18}$,
S. K. Agarwalla$^{40,\: 64}$,
J. A. Aguilar$^{12}$,
M. Ahlers$^{22}$,
J.M. Alameddine$^{23}$,
N. M. Amin$^{44}$,
K. Andeen$^{42}$,
G. Anton$^{26}$
% ...
\\
\\
% ...
$^{12}$ Universit{\'e} Libre de Bruxelles, Science Faculty CP230, B-1050 Brussels, Belgium \\
% ...
$^{17}$ Department of Physics, Loyola University Chicago, Chicago, IL 60660, USA \\
$^{18}$ Dept. of Physics and Astronomy, University of Canterbury, Private Bag 4800, Christchurch, New Zealand \\
% ...
$^{22}$ Niels Bohr Institute, University of Copenhagen, DK-2100 Copenhagen, Denmark \\
$^{23}$ Dept. of Physics, TU Dortmund University, D-44221 Dortmund, Germany \\
% ...
$^{26}$ Erlangen Centre for Astroparticle Physics, Friedrich-Alexander-Universit{\"a}t Erlangen-N{\"u}rnberg, D-91058 Erlangen, Germany \\
% ...
$^{40}$ Dept. of Physics and Wisconsin IceCube Particle Astrophysics Center, University of Wisconsin{\textendash}Madison, Madison, WI 53706, USA \\
% ...
$^{42}$ Department of Physics, Marquette University, Milwaukee, WI, 53201, USA \\
% ...
$^{44}$ Bartol Research Institute and Dept. of Physics and Astronomy, University of Delaware, Newark, DE 19716, USA \\
% ...
$^{63}$ Deutsches Elektronen-Synchrotron DESY, Platanenallee 6, 15738 Zeuthen, Germany  \\
$^{64}$ Institute of Physics, Sachivalaya Marg, Sainik School Post, Bhubaneswar 751005, India \\
% ...

\subsection*{Acknowledgements}

\noindent
The authors gratefully acknowledge the support from the following agencies and institutions:
USA {\textendash} U.S. National Science Foundation-Office of Polar Programs,
U.S. National Science Foundation-Physics Division,
U.S. National Science Foundation-EPSCoR,
Wisconsin Alumni Research Foundation,
Center for High Throughput Computing (CHTC) at the University of Wisconsin{\textendash}Madison,
Open Science Grid (OSG),
Advanced Cyberinfrastructure Coordination Ecosystem: Services {\&} Support (ACCESS),
Frontera computing project at the Texas Advanced Computing Center,
U.S. Department of Energy-National Energy Research Scientific Computing Center,
Particle astrophysics research computing center at the University of Maryland,
Institute for Cyber-Enabled Research at Michigan State University,
and Astroparticle physics computational facility at Marquette University;
Belgium {\textendash} Funds for Scientific Research (FRS-FNRS and FWO),
FWO Odysseus and Big Science programmes,
and Belgian Federal Science Policy Office (Belspo);
Germany {\textendash} Bundesministerium f{\"u}r Bildung und Forschung (BMBF),
Deutsche Forschungsgemeinschaft (DFG),
Helmholtz Alliance for Astroparticle Physics (HAP),
Initiative and Networking Fund of the Helmholtz Association,
Deutsches Elektronen Synchrotron (DESY),
and High Performance Computing cluster of the RWTH Aachen;
Sweden {\textendash} Swedish Research Council,
Swedish Polar Research Secretariat,
Swedish National Infrastructure for Computing (SNIC),
and Knut and Alice Wallenberg Foundation;
European Union {\textendash} EGI Advanced Computing for research;
Australia {\textendash} Australian Research Council;
Canada {\textendash} Natural Sciences and Engineering Research Council of Canada,
Calcul Qu{\'e}bec, Compute Ontario, Canada Foundation for Innovation, WestGrid, and Compute Canada;
Denmark {\textendash} Villum Fonden, Carlsberg Foundation, and European Commission;
New Zealand {\textendash} Marsden Fund;
Japan {\textendash} Japan Society for Promotion of Science (JSPS)
and Institute for Global Prominent Research (IGPR) of Chiba University;
Korea {\textendash} National Research Foundation of Korea (NRF);
Switzerland {\textendash} Swiss National Science Foundation (SNSF);
United Kingdom {\textendash} Department of Physics, University of Oxford.

\end{document}